# NONLINEAR DYNAMIC INTERTWINING OF RODS WITH SELF-CONTACT

**AUTHORS:**

Sachin Goyal[1(current), 2(work)] (sgoyal@whoi.edu)

N. C. Perkins[2] (ncp@umich.edu) – Corresponding Author

Christopher L. Lee[3(current), 4(work)] (christopher.lee@olin.edu)

**AFFILIATIONS:**

[1]Applied Ocean Physics and Engineering

Woods Hole Oceanographic Institution, Woods Hole MA 02543 (U.S.A.)

[2]Department of Mechanical Engineering

University of Michigan, Ann Arbor MI 48109 (U.S.A.)

[3]Department of Mechanical Engineering

Olin College, Needham, MA 02492 (U.S.A.)

[4] Lawrence Livermore National Laboratory

New Technologies Engineering Division

7000 East Ave., Livermore, CA 94550 (U.S.A.)

**ABSTRACT**

Twisted marine cables on the sea floor can form highly contorted three-dimensional loops that resemble tangles. Such tangles or 'hockles' are topologically equivalent to the plectomenes that form in supercoiled DNA molecules. The dynamic evolution of these intertwined loops is studied herein using a computational rod model that explicitly accounts for dynamic self-contact. Numerical solutions are presented for an illustrative example of a long rod subjected to increasing twist at one end. The solutions reveal the dynamic evolution of the rod from an initially straight state, through a buckled state in the approximate form of a helix, through the dynamic collapse of this helix into a near-planar loop with one site of self-contact, and the subsequent intertwining of this loop with multiple sites of self-contact. This evolution is controlled by the dynamic conversion of torsional strain energy to bending strain energy or, alternatively by the dynamic conversion of twist (Tw) to writhe (Wr).

# 1. INTRODUCTION

Cables laid upon the sea floor may form loops and tangles as illustrated in Fig. 1. The loops, sometimes referred to as *hockles*, may cause localized damage and, in the case of fiber optic cables, may also prevent signal transmission. These highly nonlinear deformations are initiated by a combination of low tension or compression (i.e. cable *slack*) and residual torsion sufficient to induce torsional buckling of the cable. Tangles evolve from a subsequent dynamic collapse of the buckled cable into highly nonlinear and intertwined configurations with self-contact.

The looped and tangled forms of marine cables are topologically equivalent to the 'plectonemic supercoiling' of long DNA molecules as illustrated in Fig. 2 (refer to Calladine et al. 2004; Goyal et al. 2005a). Figure 2 depicts a DNA molecule on three different length scales as reproduced from (Branden & Tooze 1999; Lehninger et al. 2005). The smallest length scale (far left) shows a segment of the familiar 'double-helix' which has a diameter of approximately 2 nanometers (nm). One complete helical turn is depicted here and this extends over a length of approximately 3 nm.

On an intermediate spatial scale (middle of Fig. 2), the double helix now appears as a long and slender DNA molecule that might be realized when considering tens to hundreds of helical turns (approximately tens to hundreds of nm). Two idealized 'long-length scale structures' of DNA are illustrated to the far right in Fig. 2. Here, the exceedingly long DNA molecule may contain thousands to millions of helical turns and behave as a very flexible filament with lengths ranging from micron to millimeter scales or even longer. The long-length scale curving and twisting of this flexible molecule is referred to as *supercoiling*. Two generic types of supercoils are illustrated. A *plectonemic supercoil* leads to an interwound structure where the molecule wraps upon itself with many sites of apparent 'self-contact'. By contrast, a *solenoidal supercoil* possesses no self-contact and forms a secondary helical structure resembling a coiled spring or a telephone cord.

Often with the aid of proteins, DNA must supercoil for several key reasons. First, supercoiling provides an organized means to compact the very long molecule (by as much as $10^5$) within the small confines of the cell nucleus. An unorganized compaction would hopelessly tangle the molecule and render it useless as the medium for storing genetic information. Second, supercoiling plays important roles in the transcription, regulation and repair of genes. For instance, specific regulatory proteins are known to aid or to hinder the formation of simple loops of DNA which in turn regulate gene activity; refer, for example, to (Schleif 1992) and (Semsey et al. 2005).

Like the tangling of marine cables above, the intertwining of DNA is inherently a nonlinear dynamic process controlled by structural properties (e.g., elasticity) and applied forces (e.g., protein interactions). Rod theory provides a useful framework to explore the dynamics of intertwining of long filament-like structures such as cables and DNA molecules, as described, for example in (Goyal 2006). The mechanics of intertwining immediately invokes formulations for self-contact in rod theory which remain a significant challenge as emphasized recently in (Chouaieb et al. 2006; van der Heijden et al. 2006).

The inclusion of self-contact in *equilibrium* formulations of rod theory has been treated in (Coleman et al. 2000; Gonzalez et al. 2002; Schuricht & von der Mosel 2003; van der Heijden et al. 2003; Coleman & Swigon 2004; Chouaieb et al. 2006; van der Heijden et al. 2006). In particular, Chouaieb et al. (2006) evaluate helical equilibria where self-contact is accounted for by imposing bounds on helical curvature and torsion. The formation of self-contact in the equilibria generated from 'closed' or 'circular' rods (e.g., representative of DNA plasmids) is examined in (Coleman et al. 2000; Coleman & Swigon 2004) using numerical energy minimization. The mathematical existence of such solutions is deduced in (Gonzalez et al. 2002) by careful formulation of the geometric excluded volume constraint on self-intersection. The excluded volume constraint is formulated in terms rod of centerline curvature in (Schuricht & von der Mosel 2003) and appended via Lagrange multiplier to the Euler-Lagrange equation for rod equilibrium with self-contact. The analysis of 'open' rods (e.g., rods that do not close upon themselves) requires consideration of two boundary conditions through which loads may also be applied. A numerical study of the self-contacting equilibria of clamped-clamped rods reveals the bifurcations generated by varying compression and twist applied at the boundaries (van der Heijden et al. 2003). A recent extension (van der Heijden et al. 2006) considers cases where the rod is constrained to lie on the surface of a cylinder. Open questions regarding the analysis of rods with self-contact are emphasized in (van der Heijden et al. 2006) by the lament "*We are still far from understanding analytically the solutions of the Euler-Lagrange equations for general contact situations. Even if we limit ourselves to global minimizers of an appropriate energy functional, we can prove little about the form of solutions as soon as contact is taken into account.*"

In contrast to the equilibrium formulations above, very few *dynamical* formulations of rod theory have been proposed that incorporate self-contact. Nevertheless, such formulations enable one to explore the dynamic evolution of self-contacting states and possible dynamic transitions between them. For instance, the slow twisting of the filament treated in (Goyal et al. 2003b) ultimately induces a sudden dynamic collapse of an intermediate helical loop into an intertwined form. An approximate dynamical formulation is also presented in (Klapper 1996) where inertial effects are ignored in favor of dissipation and stiffness effects.

In this paper, we revisit the slow twisting of a filament (Goyal et al. 2003b) with the objective to develop a fundamental understanding of the dynamic evolution of its intertwined states. In particular, we describe how intertwined states result from a sudden collapse of helically-looped states through a rapid conversion of torsional to bending strain energy. The remainder of this paper is organized as follows. Sections 2 and 3 summarize a computational dynamic rod model that incorporates self-contact (Goyal 2006). Section 4 presents an illustrative example of a non-homogeneous rod subject to pure torsion. Results highlight the dynamic evolution from straight to looped to intertwined states following a dramatic collapse to self-contact. We close in Section 5 with conclusions.

## 2. COMPUTATIONAL ROD MODEL – A SUMMARY

The rod segment illustrated in Fig. 3 is a thin (1-dimensional) element that may undergo two-axis bending and torsion in forming a three-dimensional space curve. This curve represents the rod centerline which, in the context of double-stranded DNA, represents the helical axis of the duplex. We develop the dynamical model by employing the classical approximations of Kirchhoff and Clebsch (Love 1944) as detailed in (Goyal 2006). A summary is provided here.

### 2.1 Rod Kinematics, Constitutive Law, and Energy

Consider the infinitesimal element of a Kirchhoff rod shown in Fig. 3. The three-dimensional curve $R(s,t)$ formed by the centerline is parameterized by the arc length coordinate $s$ and time $t$. The body-fixed frame $\{a_i\}$ at each cross-section is employed to describe the orientation of the cross-section with respect to the inertial frame $\{e_i\}$. The angular velocity $\omega(s,t)$ of the cross-section is defined as the rotation of the body-fixed frame $\{a_i\}$ per unit time relative to the inertial frame $\{e_i\}$ and satisfies

$$\left(\frac{\partial a_i}{\partial t}\right)_{\{e_i\}} = \omega \times a_i, \tag{1}$$

where the subscript specifies the reference frame relative to which the derivative has been taken. We also define a 'curvature and twist vector' $\kappa(s,t)$ as the rotation of the body-fixed frame $\{a_i\}$ per unit arc length relative to the inertial frame $\{e_i\}$ which satisfies

$$\left(\frac{\partial a_i}{\partial s}\right)_{\{e_i\}} = \kappa \times a_i. \tag{2}$$

In a stress-free state, the rod conforms to its natural geometry defined by $\kappa_0(s)$. The difference $\{\kappa(s,t) - \kappa_0(s)\}$ results in an internal moment $q(s,t)$ at each cross-section of the rod. The relationship between the change in curvature/twist $\{\kappa(s,t) - \kappa_0(s)\}$ and the restoring moment $q(s,t)$ is governed by a constitutive law for bending and torsion. While many generalizations of the constitutive law are discussed in (Goyal 2006), in this study we employ the linear elastic law

$$q(s,t) = B(s)(\kappa(s,t) - \kappa_0(s)) \tag{3}$$

where $B(s)$ is a positive definite stiffness tensor that is a prescribed function of position $s$. The resulting strain energy density is

$$S_e(s,t) = \frac{1}{2}(\kappa(s,t) - \kappa_0(s))^T B(s)(\kappa(s,t) - \kappa_0(s)). \tag{4}$$

We further employ a diagonalized form of $B(s)$ by choosing $\{a_i\}$ to coincide with the 'principal torsion-flexure axes' of the cross-section (Love 1944). In particular, $a_1$ and $a_2$ are in the plane of the cross-section and are aligned with the principal flexure axes while $a_3$ is normal to the cross-section and coincides with the tangent $\hat{t}$. The resulting diagonal form of the stiffness tensor $B(s)$ is

$$B(s) = \begin{bmatrix} A_1(s) & 0 & 0 \\ 0 & A_2(s) & 0 \\ 0 & 0 & C(s) \end{bmatrix}, \qquad (5)$$

where $A_1(s)$ and $A_2(s)$ are bending stiffnesses about the principal flexure axes along $a_1$ and $a_2$ respectively, and $C(s)$ is the torsional stiffness about principal torsional or 'tangent' axis $a_3$. Furthermore, in the results that follow, the rod is assumed to be isotropic[1] but non-homogeneous (i.e. $A_1(s) = A_2(s) = A(s)$). The stress-distribution at any cross-section not only results in a net internal moment $q(s,t)$, but also net tensile and shear forces $f(s,t)$.

The kinetic energy of the rod depends upon the centerline velocity $v(s,t)$ and the cross-section angular velocity $\omega(s,t)$. Let $m(s)$ denote the mass of the rod per unit arc length and $I(s)$ denote the tensor of principal mass moments of inertia per unit arc length. Then the rod kinetic energy density is

$$K_e(s,t) = \frac{1}{2}\omega(s,t)^T I(s)\omega(s,t) + \frac{1}{2}v(s,t)^T m(s)v(s,t). \qquad (6)$$

---

[1] The rod is assumed to have circular cross section in this study with axi-symmetric bending stiffness.

We choose the vectors $v(s,t)$, $\omega(s,t)$, $\kappa(s,t)$ and $f(s,t)$ as four unknown field variables in the formulation below. The kinematical quantities $\kappa(s,t)$, $\omega(s,t)$ and $v(s,t)$ can be readily integrated to compute the rod configuration $R(s,t)$ and the cross-section orientation as given by $\{a_i(s,t)\}$; refer to Fig. 3 and to (Goyal 2006).

Depending upon the application, the rod may also interact with numerous external field forces including those produced by gravity, a surrounding fluid medium, electrostatic forces, contact with other bodies or with the rod itself, , etc. The resultant of these external forces and moments per unit length is denoted by $F(s,t,...)$ and $Q(s,t,...)$, respectively. In general, these quantities may be functionally-dependent on the kinematical quantities $\kappa(s,t)$, $\omega(s,t)$ and $v(s,t)$ in addition to the rod configuration $R(s,t)$.

We next specify the four field equations required to solve for the four vector unknowns $\{v, \omega, \kappa, f\}$. In the field equations, we employ partial derivatives of all quantities relative to the body-fixed frame $\{a_i\}$ and recall the following relations to the partial derivatives relative to the inertial frame for a vector quantity $\upsilon$ (Greenwood 1988):

$$\left(\frac{\partial \upsilon}{\partial t}\right)_{\{a_i\}} = \left(\frac{\partial \upsilon}{\partial t}\right)_{\{e_i\}} - \omega \times \upsilon \quad \text{and} \quad \left(\frac{\partial \upsilon}{\partial s}\right)_{\{a_i\}} = \left(\frac{\partial \upsilon}{\partial s}\right)_{\{e_i\}} - \kappa \times \upsilon, \tag{7}$$

For notational convenience, we drop the subscript for the body-fixed frame from this point forward.

## 2.2 Equations of Motion

The balance law for linear momentum of the infinitesimal element shown in Fig. 3 becomes

$$\frac{\partial f}{\partial s} + \kappa \times f = m\left(\frac{\partial v}{\partial t} + \omega \times v\right) - F \qquad (8)$$

and that for angular momentum becomes

$$\frac{\partial q}{\partial s} + \kappa \times q = I\frac{\partial \omega}{\partial t} + \omega \times I\omega + f \times \hat{t} - Q, \qquad (9)$$

Here, $\hat{t}(s,t)$ is the unit tangent vector along the centerline (directed towards increasing arc length $s$) and the internal moment $q(s,t) = B(s)(\kappa(s,t) - \kappa_0(s))$ upon substitution of the constitutive law Eq. (3).

## 2.3 Constraints and Summary

The above formulation is completed with the addition of two vector constraints. The first enforces *inextensibility* and *unshearability* which take the form

$$\frac{\partial v}{\partial s} + \kappa \times v = \omega \times \hat{t}. \qquad (10)$$

The second follows from continuity requirements for $\omega$ and $\kappa$ in the form of the *compatibility* constraint

$$\frac{\partial \omega}{\partial s} + \kappa \times \omega = \frac{\partial \kappa}{\partial t}. \tag{11}$$

Detailed derivations of these constraints are provided in (Goyal 2006).

The four vector equations Eq. (8-11) in the four vector unknowns $\{v, \omega, \kappa, f\}$ result in a $12^{th}$ order system of nonlinear partial differential equations in space and time. They are compactly written as

$$M(Y,s,t)\frac{\partial Y}{\partial t} + K(Y,s,t)\frac{\partial Y}{\partial s} + F(Y,s,t) = 0 \tag{12}$$

where $Y(s,t) = \{v, \omega, \kappa, f\}$ and the operators $M$, $K$ and $F$ are described in ( Goyal et al. 2005b; Goyal 2006;). These equations are not integrable in general and thus we pursue a numerical solution as detailed in (Goyal et al. 2005b; Goyal 2006). In particular, we discretize the equations above by employing a finite difference algorithm using the generalized-α method (Chung & Hulbert 1993) in both space and time. Doing so yields a method that is unconditionally stable and second-order accurate. A single numerical parameter can be varied to control maximum numerical dissipation. The difference equations so obtained are implicit and their solution must satisfy the rod boundary conditions. The boundary conditions are satisfied using a shooting method in conjunction with Newton-Raphson iteration. In addition, this formulation also incorporates the forces generated by self-contact which, being central to the objective of this paper, we describe in some detail below.

## 3. NUMERICAL FORMULATION OF DYNAMIC SELF-CONTACT

A numerical formulation of self-contact begins with first determining the likely sites where self-contact exists or will soon occur. An efficient search strategy for these sites (Goyal 2006) is as follows. Consider two remote segments of the discretized rod that are approaching contact as shown in Fig. 4. The lower segment contains three spatial grid points denoted as 1, 2 and 3 while the upper segment contains one grid point denoted as 4. Grid point 4 is likely to interact with the grid point 2 as the two segments approach each other. We introduce a *screening aperture* of angle $\theta$ formed by a pair of conical surfaces centered at each grid point (illustrated at grid point 2 in Fig. 4). We use this aperture to efficiently search for only those points that may potentially interact through self-contact. This aperture specifically excludes non-physical 'contact' forces between nearby grid points on the same segment (such as 1, 2 and 3 in lower segment). The aperture reduces to the plane of the rod cross-section as $\theta \rightarrow 0°$, and it expands to the entire space as $\theta \rightarrow 180°$.

During simulation, the separation $d$ between each pair of grid points is measured. A repulsive (contact) force is introduced between these grid points only if two conditions are met: 1) the distance $d$ is less than a specified tolerance, and 2) the two grid points lie within each other's *screening aperture*. This search strategy ensures that the contact forces are approximately normal to the rod surfaces and also allows for sliding contact. The interaction force can in general be a function of $d$ and $\dot{d}$ (the approach speed) and it is included in the balance of linear momentum Eq. (8) through the distributed force term $F$. Example interaction laws that can be employed include (attractive-repulsive) Lennard-Jones type (refer to, for example (Schlick et al. 1994b)), (screened repulsion) Debye-Huckle type (refer to, for example (Schlick et al. 1994a)), general inverse-power laws (refer to, for example (Klapper 1996)), and idealized contact laws for two solids (refer to, for example (van der Heijden et al. 2003) and (Coleman et al. 2000)). In the specific case of DNA, one might introduce a fictitious charged and cylindrical surface that circumscribes the molecule to capture the repulsive effects of the negatively charged backbone.

## 4. RESULTS

The computational model above is used to explore the dynamic evolution of an intertwined state induced by slowly increasing the twist applied to one end of an elastic rod. The numerical solutions reveals three major behaviors: 1) the torsional buckling of an initially straight rod into the approximate shape of a helix, 2) the dramatic collapse of this helix to a near-planar loop with self-contact at a single point, and 3) the subsequent intertwining of the loop with multiple sites of self-contact.

**4.1 Illustrative Example**

Figure 5 defines an illustrative example which consists of an initially straight, linearly elastic rod subjected to monotonically increasing twist at the right end at $s = 0$. This end cannot move and it is otherwise constrained in rotation (no rotation about the principal axes $a_1$ and $a_2$). The left end at $s = L$ is fully restrained in rotation and cannot translate in the transverse ($a_1$-$a_2$) plane. This end, however, may translate along the $e_2$ axis.

The material and geometric parameters that define the example are listed in Table 1 together with basic discretization parameters used in the numerical algorithm; refer to (Goyal et al. 2005b; Goyal 2006) for a complete description of the numerical parameters. The example rod has a circular cross-section which varies along its length. In particular, the central portion of the rod (middle 25%) is necked down to a smaller diameter that is 10% smaller than the end regions. We have chosen this non-homogenous rod to illustrate both the generality of the computational model as well as to promote torsional buckling and subsequent intertwining within the ('softer') central portion. The small 10% reduction in the diameter produces a significant ($\approx 35\%$) reduction in torsional stiffness ($C(s) = GJ_3$) and bending stiffness ($A(s) = EJ_{1,2}$) in the central portion.

As a representative law for self-contact, we choose for this example the following repulsive force

$$|F_{contact}| = \rho_c A_c \left( \frac{k_1}{(d - 0.5D)^{k_2}} + \frac{k_3}{d} \dot{d}|\dot{d}|^{k_4} \right), \tag{13}$$

with example parameters: $k_1 = 10^{-7} \text{m}^4/\text{s}^2$, $k_2 = 3$, $k_3 = 10^{-6}$ and $k_4 = 1$. This contact law is one of many possible that capture both nonlinear repulsion and dissipation. The results that follow are rather insensitive to changes in the specific parameter values selected for this example.

In addition to the contact law above, the only other body force considered is a dissipative force. As one example, we introduce the viscous drag imparted by a surrounding fluid environment in the form of the standard Morison drag law (Morison et al. 1950). This distributed drag, which manifests itself in the balance of linear momentum Eq. (8) through the distributed force term $F$, is computed as (Goyal et al. 2005b):

$$F_{drag} = -\frac{1}{2} \rho_f D \{ C_n |v \times \hat{t}| \hat{t} \times (v \times \hat{t}) + \pi C_t (v \cdot \hat{t}) v \cdot \hat{t} | \hat{t} \}, \tag{14}$$

Here, $C_n$ is the normal (form) drag coefficient, $C_t$ is the tangential (skin friction) drag coefficient, and $\rho_f$ is fluid density. Example values of these parameters are reported in Table 1.

**4.2 Evolution of Self-Contact and Intertwining**

By increasing the rotation (twist) slowly at the right end, the internal torque eventually reaches the bifurcation condition associated with the classical torsional buckling of a straight rod (Zachmann 1979). This rotation is generated by prescribing the angular velocity component $\omega_3$ at the right end as shown in Fig. 6 (not to scale). In addition, the left end is allowed to translate freely during the first 30 seconds and is then held fixed to control what would otherwise be an exceedingly rapid collapse to self-contact as described in the following.

As the right end is initially twisted by a modest amount, the rod remains straight. There is an abrupt change however when the twist reaches the bifurcation value associated with the Zachmann buckling condition (Zachmann 1979) and the straight (trivial) configuration becomes unstable. This occurs at approximately 16 seconds in this example. The computational model captures this initial instability as well as the subsequent nonlinear motion that leads to loop formation and ultimately to intertwining.

Figure 7 illustrates four representative snap-shots during the dynamic evolution of an intertwined state. The geometry just after initial buckling is approximately helical as can be observed in the uppermost snap-shot (20 seconds). Notice that the rod centerline appears to make a single helical turn as predicted from the fundamental buckling mode of the linearized theory (Zachmann 1979). The superimposed black stripe records the computed twist distribution of the rod for this state which exhibits nearly four complete turns; refer to the discussion of twist and major topological transitions below. As this twist is increased, the rod continuously deforms into a larger diameter helical loop and the left end slides substantially to the right as shown by the second image (25 seconds).

Upon greater twist, the left end continues to slide towards the right end and the helical loop continues to rotate out of the plane of this figure. Eventually the loop undergoes a secondary bifurcation followed by a rapid dynamic collapses into self-contact in forming a nearly planar loop. The collapsed loop is shown by the third image (which occurs at approximately 29 seconds).

The dynamic collapse can be anticipated from stability analyses of the equilibrium forms of a rod under similar loading conditions; refer to (Lu & Perkins 1994) and studies cited therein. The snap-shot at 25 seconds shows the three-dimensional shape of the rod just prior to dynamic collapse. Here, the apex of the loop has rotated approximately 90° about the vertical ($e_1$) axis so that the tangent at the apex is now orthogonal to the loading ($e_2$) axis. This was the noted secondary bifurcation condition in (Lu & Perkins 1994) at which the three-dimensional equilibrium form loses stability.

The collapsed loop, however, is very sensitive to the increasing twist and rapidly continues to rotate about the vertical ($e_1$) axis leading to intertwined forms with multiple sites of self-contact. A snapshot of a fully intertwined loop is illustrated at the bottom of Fig. 7 (32 seconds). The strain energy density (color scale in Fig. 7) reveals that the strain energy becomes highly localized to the apex of the intertwined loop where the curvature is greatest. The decomposition of this strain energy into bending and torsional components provides significant insight into the dynamic evolution of an intertwined state as discussed next.

**4.3 Energetic Transitions**

Figure 8 summarizes the energetics of this process by illustrating how the bending and torsional strain energy components contribute to the total strain energy. Starting at time zero, the initially straight rod remains straight and the applied twist simply increases the torsional strain energy. This elementary, pure-twisting of the straight rod ceases at approximately 18 seconds with the first bifurcation due to torsional buckling (Zachmann 1979). The torsional strain energy achieves its maximum at this state and immediately thereafter the rod buckles into a three-dimensional form resembling a shallow helix (a). This transition is accompanied by a conversion of torsional to bending strain energy. This conversion is dynamic and markedly increases as the rod is twisted further while developing a distinctive loop (b). The apex of this loop rotates further out of plane during this stage. Just prior to 29 seconds the apex becomes orthogonal to the loading axis (original axis of the straight rod) which marks the secondary bifurcation (Lu & Perkins 1994) that generates an extremely fast dynamic collapse to self-contact. The resulting loop with self-contact is nearly planar (c). During this secondary bifurcation, the rod loses both torsional and bending strain energies until self-contact and, thereafter intertwining begins. As intertwining advances (d), the torsional strain energy continues to decrease while the bending strain energy increases once more. In addition, the bending strain energy becomes localized to the apex of the loop due to the significant and increasing curvature developed there; refer also to snapshot at 32 seconds in Fig. 7. In the case of DNA forming plectonemes, such localized strain energy might possibly be the forerunner of the nonlinear 'kinking' of the molecule as proposed recently (Wiggins et al. 2005). Figure 8 also illustrates the total strain energy and the work done by twisting the right boundary. The energy difference between the work done and the total strain energy derives from the significant kinetic energy during this process as well as the dissipation developed from the included fluid drag.

**4.4 Topological Transitions**

It is interesting to observe that the topological changes for the example rod above are also exhibited by DNA during supercoiling. As discussed in (Calladine et al. 2004), the above conversion of torsional strain energy to bending strain energy for DNA is more frequently described topologically as the conversion of *twist* to *writhe*. We explore this conversion in the above example after briefly reviewing the definitions for twist and writhe.

Twist (Tw) is a kinematical quantity representing the total number of twisted turns along the rod centerline as computed from

$$Tw = \frac{1}{2\pi} \int_0^{L_C} (\kappa \cdot \hat{t}) ds \qquad (15)$$

Writhe (Wr) is defined as the average number of cross-overs of the rod centerline when observed over all possible views of the rod (Calladine et al. 2004). For our initially straight configuration, Wr = 0. At the first self-contact shown by the snapshot at 29 seconds in Fig. 7, Wr = 1. The writhe then continues to increase to Wr = 2 for the intertwined state at 32 seconds in Fig. 7. The writhe is purely a function of the space curve defining the rod centerline and it may also be positive or negative depending on whether the crossing is right-handed or left-handed (Calladine et al. 2004). In our illustrative example, the sum Tw + Wr equals the number of rotations of the right boundary and this sum is called the *Linking number* Lk[2]. Refer to Fuller (Fuller 1971) and White (White 1969) for the proof of conservation of the Linking number (Lk).

In our example, the initial twisting phase rapidly increases Lk from 0 to approximately 4, all in the form of twist, prior to the first bifurcation (torsional buckling) as illustrated in Fig. 9. An additional increase in Lk (end rotation) of less than ½ (turn) produces all of the sudden transitions noted above. Following the first bifurcation, Wr increases from 0 to 1 at self-contact (29 seconds) and Tw correspondingly reduces so that the sum Wr + Tw remains equal to Lk. Following the first self-contact, the loop continues to rotate as it intertwines. In doing so, every half rotation of the loop establishes an additional contact site thereby increasing Wr by 1 and reducing Tw by 1. At 32 seconds, Wr is slightly larger than 2. Thus, we observe two crossovers in any three orthogonal views of the snapshot at 32 seconds shown in Fig. 7. There is a compensatory loss in Tw as shown in Fig. 9.

It should also be noted if self-contact is ignored, as has often been done in some prior studies of the looping of rods, the numerical solution for the rod may allow it to artificially 'cut through itself' leading to entirely different and non-physical results. Following each 'cut', both Wr and Lk are reduced discontinuously by 2. Examples of this readily follow from the present computational formulation by simply eliminating the contact force. However, doing so leads to non-physical discontinuous changes in Wr and Lk following artificial 'cuts' through the rod. Thus, modeling self-contact is fundamentally necessary when one endeavors to understand the pathway(s) leading to the intertwined loops.

---

[2] This is not true, in general, for other boundary conditions that allow rotations about the other axes (i.e., cases where $\omega_{1,2} \neq 0$ at the boundaries).

## 5. CONCLUSIONS

This paper summarizes a computational rod model that captures the dynamical evolution of intertwined loops in rods under torsion. A major feature is the explicit formulation of dynamic self-contact. An illustrative example is selected which reveals a fundamental understanding of how loops first form, then collapse, and then intertwine. This knowledge may also promote an understanding of how long cables form 'hockles' and how DNA molecules form plectonemic supercoils.

Numerical simulations reveal that an originally straight rod undergoes two bifurcations in succession as twist is added. The first bifurcation is elementary and occurs at the (Zachmann) buckling condition where the trivial equilibrium becomes unstable and the rod buckles into the approximate shape of a shallow helix. Upon increasing twist, this helix grows in amplitude to form a distinctive loop. In doing so, the apex of this loop continues to rotate towards the out-of-plane direction. When the apex ultimately becomes orthogonal to the loading axis (axis of the original straight rod), the loop experiences a secondary bifurcation and a sudden dynamic collapse into a near-planar loop with self-contact. As twist is again added, the near-planar loop rotates upon itself becoming intertwined with multiple sites of self-contact. The energetics leading to the intertwined form confirm the large exchange of torsional strain energy for bending strain energy which becomes increasingly localized to the apex of the loop. These transitions parallel the dynamic conversion of twist (Tw) to writhe (Wr) during this process.

**ACKNOWLEDGEMENTS**

The authors (NCP and SG) gratefully acknowledge the research support provided by the U.S. Office of Naval Research (grant N00014-00-1-0001), the Lawrence Livermore National Laboratories (grant B 531085), and the U. S. National Science Foundation (grants CMS 0439574 and CMS 0510266). The author (CLL) performed work under the auspices of the U.S. Department of Energy by the University of California, Lawrence Livermore National Laboratory under contract No.W-7405-ENG-48.

| quantity | units (SI) | value/ formula |
|---|---|---|
| Young's Modulus, $E$ | Pa | $1.25 \times 10^7$ |
| Shear Modulus, $G$ | Pa | $5.0 \times 10^6$ |
| Diameter, $D$ | m | See Fig. 5 |
| Length, $L_c$ | m | 1.0 |
| Rod Density, $\rho_c$ | Kg/m$^3$ | 1500 |
| Fluid Density, $\rho_f$ | Kg/m$^3$ | 1000 |
| Normal Drag Coefficient $C_n$ | - | 0.1 |
| Tangential Drag Coefficient $C_t$ | - | 0.01 |
| Temporal Step, $\Delta t$ | s | 0.1 |
| Spatial Step, $\Delta s$ | m | 0.001 |
| Cross-section Area | m$^2$ | $A_c = \dfrac{\pi D^2}{4}$ |
| Mass/ length | Kg/m | $m = \rho_c A_c$ |
| Area Moments of Inertia (bending) | m$^4$ | $J_{1,2} = \dfrac{A_c D^2}{16}$ |
| Area Moment of Inertia (torsion) | m$^4$ | $J_3 = \dfrac{A_c D^2}{8}$ |
| Mass Moment of Inertia/ length | Kg-m | $I = \rho_c J$ |

**Table 1:** Example rod properties and simulation parameters.

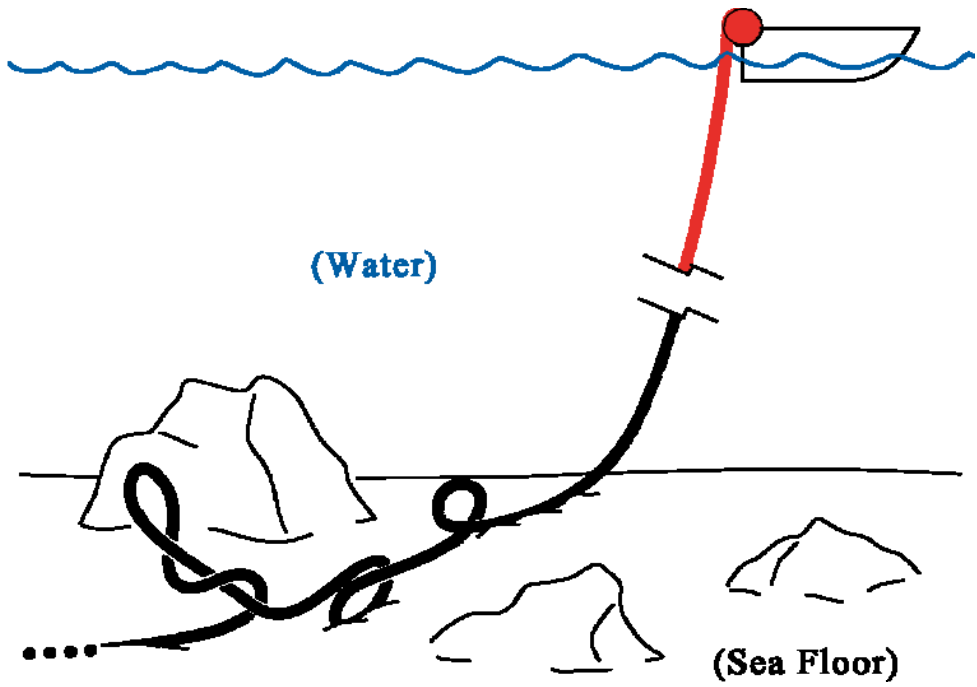

**Figure 1:** Low tension cable forming loops and (intertwined) tangles on the sea floor.

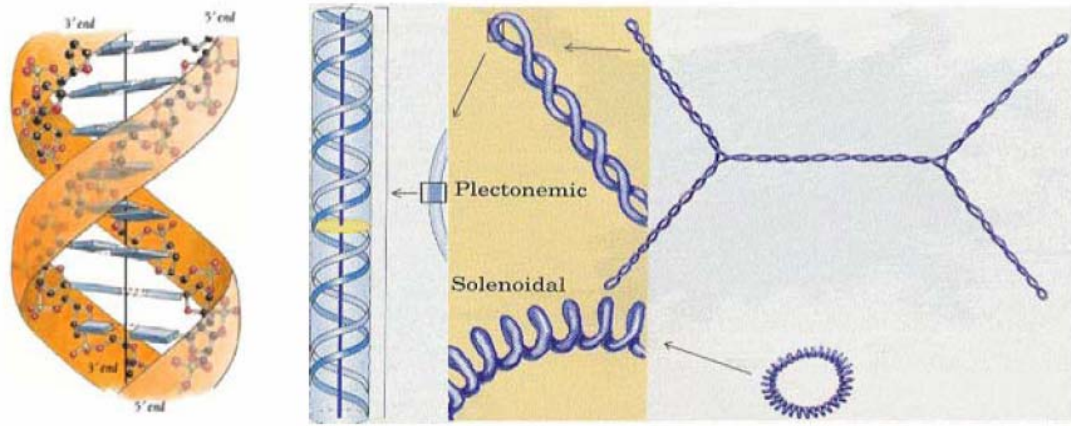

**Figure 2:** DNA shown on three length scales. Smallest scale (left) shows a single helical repeat of the double-helix structure (sugar-phosphate chains and base-pairs). Intermediate scale (middle) suggests how many consecutive helical repeats form the very long and slender DNA molecule. Largest scale (right) shows how the molecule ultimately curves and twists in forming supercoils (plectonemic or solenoidal). (Courtesy: (Branden & Tooze 1999) and (Lehninger et al. 2005)).

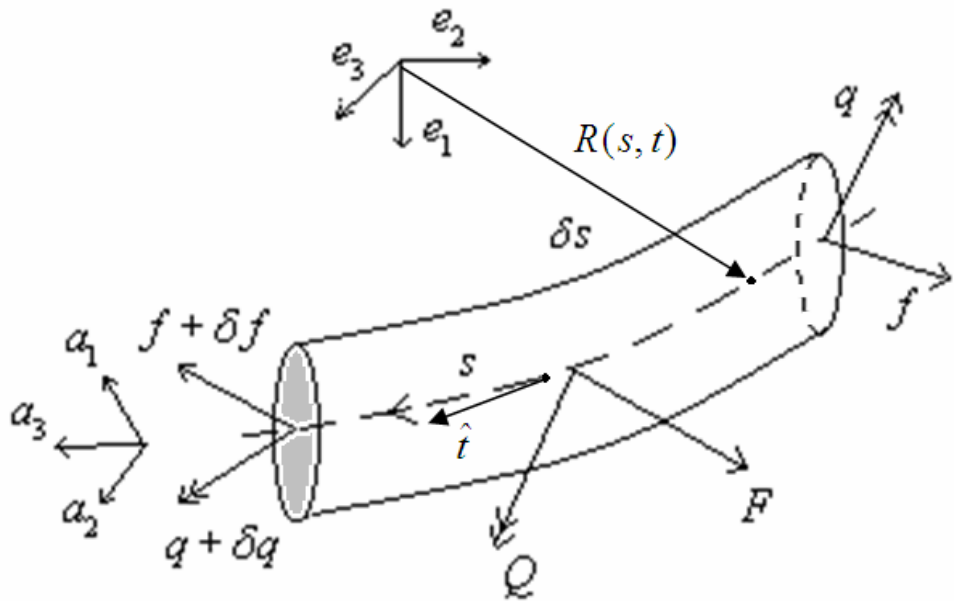

**Figure 3:** Free body diagram of an infinitesimal element of a Kirchhoff rod.

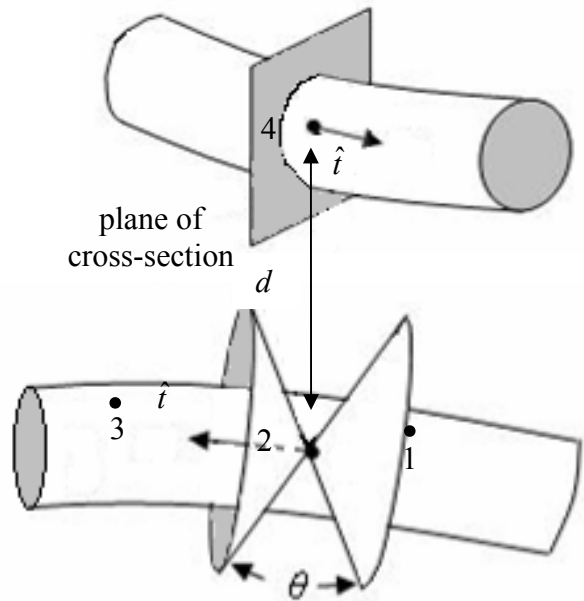

**Figure 4:** Two remote segments of a rod approaching contact. A *screening aperture* is defined by a pair of conical surfaces constructed at each grid point. This aperture leads to an efficient numerical search for regions of self-contact.

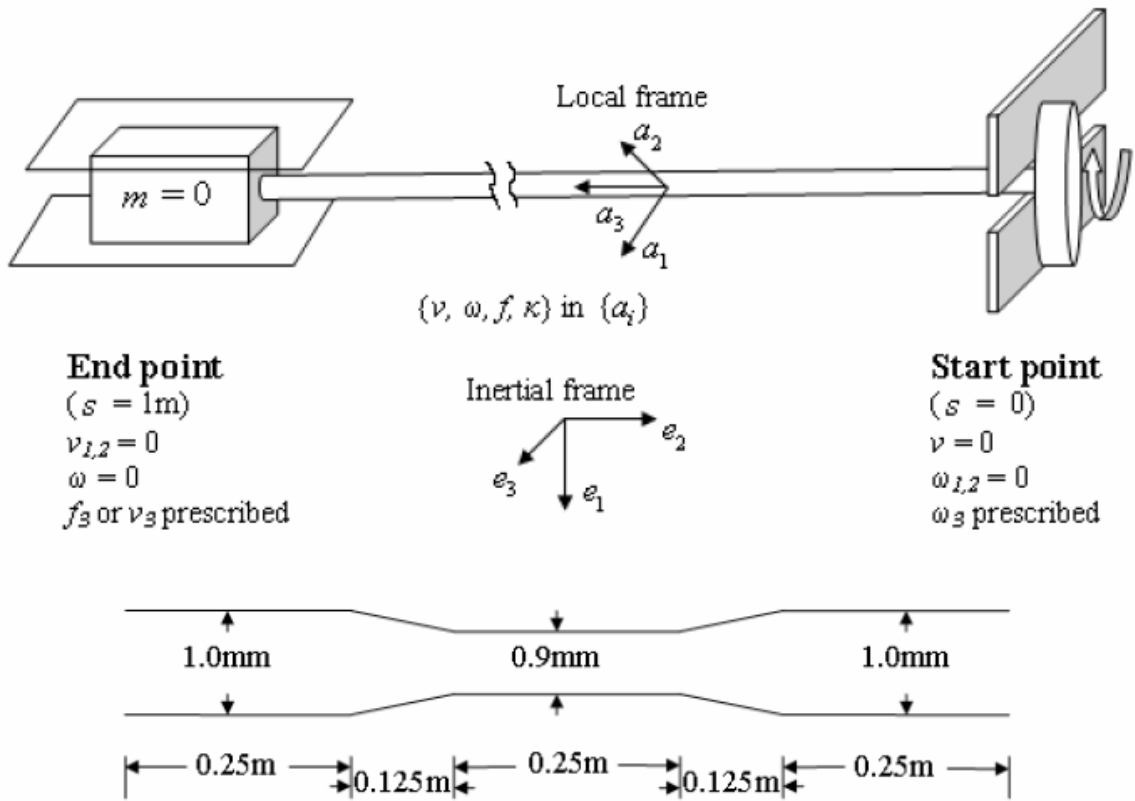

**Figure 5:** A non-homogenous rod subject to slowly increasing twist created by rotating the right end about the $e_2$ (loading) axis The right end is otherwise restrained in rotation and translation. The left end is fully restrained in rotation and translation except that it is free to slide along the loading axis.

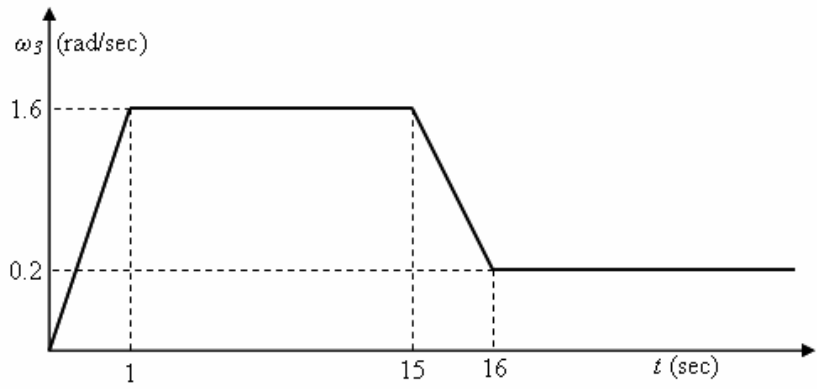

**Figure 6:** Prescribed angular (twist) velocity at the right end. (Note: not to scale).

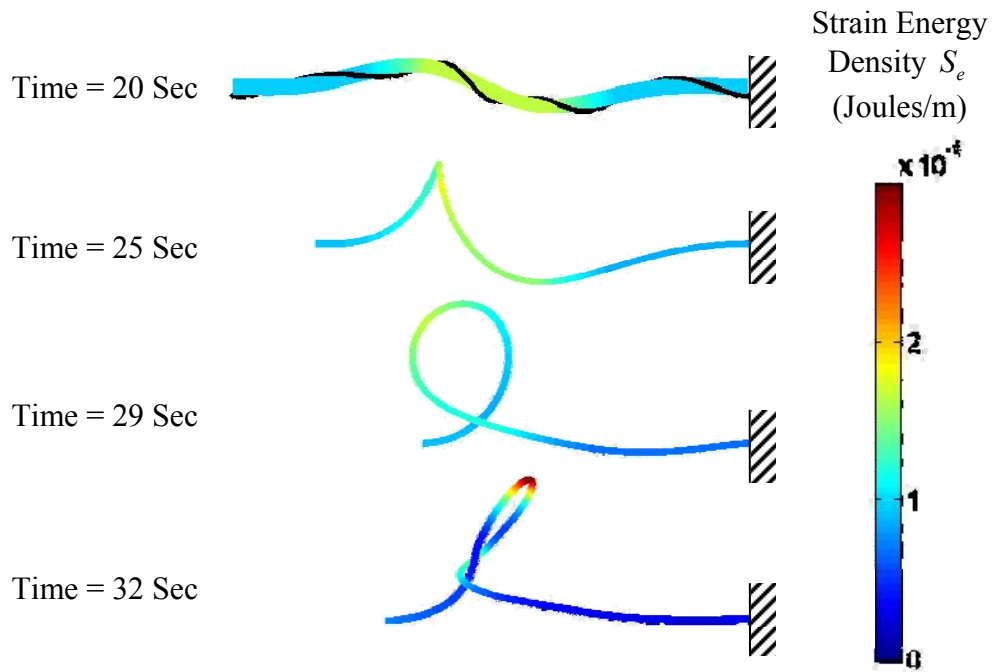

**Figure 7:** Snap-shots at selected times during the transition from a buckled helical form (Time=20 sec.) to an intertwined form (Time=32 sec.). Black stripe superimposed on the first form illustrates the twist distribution. Color indicates strain energy density.

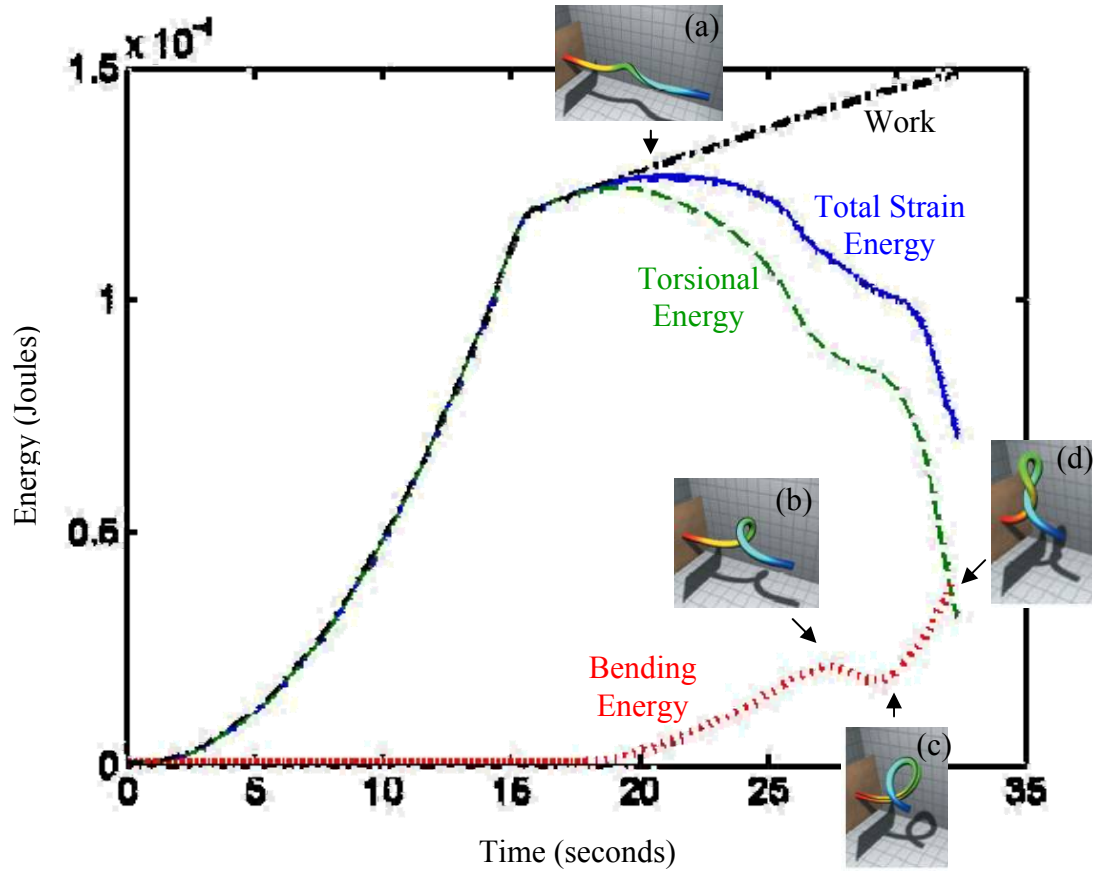

**Figure 8:** The bending, torsional, and total strain energy during the dynamic evolution of an intertwined state. The work done by the applied twist is also reported.

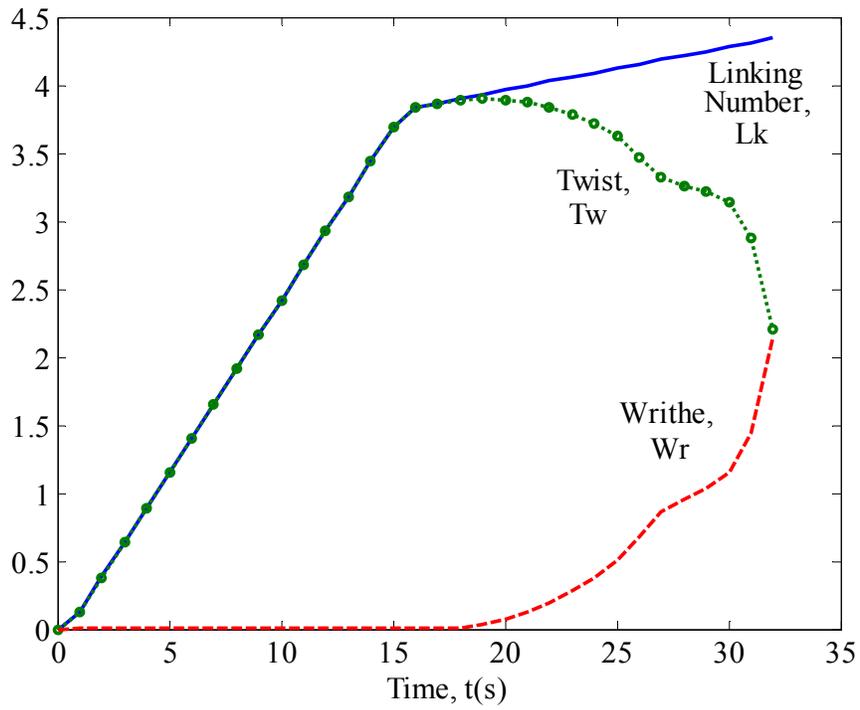

**Figure 9:** Conversion of twist (Tw) to writhe (Wr) during loop formation and intertwining. The linking number Lk = Tw + Wr is equivalent to the number of turns prescribed at the right end of the rod in this example.